\documentclass[8.5pt,twoside,twocolumn]{article}
\usepackage[super,sort&compress,comma]{natbib} 
\usepackage{mhchem}
\usepackage{times}
\usepackage{sectsty}
\usepackage{balance} 

\usepackage{graphicx} 
\usepackage{lastpage}
\usepackage[format=plain,justification=raggedright,singlelinecheck=false,font=small,labelfont=bf,labelsep=space]{caption} 
\usepackage{fancyhdr}
\usepackage{amssymb,amsmath,MnSymbol,ifsym,latexsym,bm}
\usepackage[Euler]{upgreek}
\usepackage{bm}
\usepackage{color}

\begin{document}

\thispagestyle{plain}
\fancypagestyle{plain}{
\fancyhead[L]{\includegraphics[height=8pt]{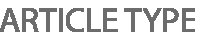}}
\fancyhead[C]{\hspace{-1cm}\includegraphics[height=20pt]{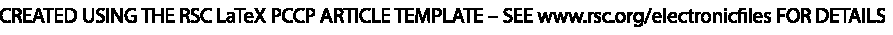}}
\fancyhead[R]{\includegraphics[height=10pt]{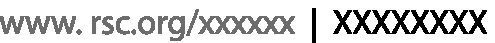}\vspace{-0.2cm}}
\renewcommand{\headrulewidth}{1pt}}
\renewcommand{\thefootnote}{\fnsymbol{footnote}}
\renewcommand\footnoterule{\vspace*{1pt}%
\hrule width 3.4in height 0.4pt \vspace*{5pt}} 
\setcounter{secnumdepth}{5}

\makeatletter 
\def\subsubsection{\@startsection{subsubsection}{3}{10pt}{-1.25ex plus -1ex minus -.1ex}{0ex plus 0ex}{\normalsize\bf}} 
\def\paragraph{\@startsection{paragraph}{4}{10pt}{-1.25ex plus -1ex minus -.1ex}{0ex plus 0ex}{\normalsize\textit}} 
\renewcommand\@biblabel[1]{#1}            
\renewcommand\@makefntext[1]%
{\noindent\makebox[0pt][r]{\@thefnmark\,}#1}
\makeatother 
\renewcommand{\figurename}{\small{Fig.}~}
\sectionfont{\large}
\subsectionfont{\normalsize} 

\fancyfoot{}
\fancyfoot[LO,RE]{\vspace{-7pt}\includegraphics[height=9pt]{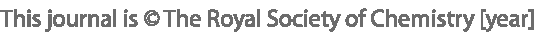}}
\fancyfoot[CO]{\vspace{-7.2pt}\hspace{12.2cm}\includegraphics{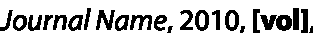}}
\fancyfoot[CE]{\vspace{-7.5pt}\hspace{-13.5cm}\includegraphics{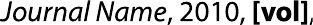}}
\fancyfoot[RO]{\footnotesize{\sffamily{1--\pageref{LastPage} ~\textbar  \hspace{2pt}\thepage}}}
\fancyfoot[LE]{\footnotesize{\sffamily{\thepage~\textbar\hspace{3.45cm} 1--\pageref{LastPage}}}}
\fancyhead{}
\renewcommand{\headrulewidth}{1pt} 
\renewcommand{\footrulewidth}{1pt}
\setlength{\arrayrulewidth}{1pt}
\setlength{\columnsep}{6.5mm}
\setlength\bibsep{1pt}

\twocolumn[
  \begin{@twocolumnfalse}
\noindent\LARGE{\textbf{Extensional viscosity of copper nanowire suspensions in an aqueous polymer solution}}
\vspace{0.6cm}

\noindent\large{\textbf{Amarin G. M$^c$Donnell,\textit{$^{a}$} Naveen N. Jason,\textit{$^{b}$} Leslie Y. Yeo,\textit{$^{c}$} James R. Friend,\textit{$^{d}$} Wenlong Cheng,\textit{$^{b}$} and Ranganathan Prabhakar$^{\ast}$\textit{$^{a}$}}}\vspace{0.5cm}

\noindent\textit{\small{\textbf{Received Xth XXXXXXXXXX 20XX, Accepted Xth XXXXXXXXX 20XX\newline
First published on the web Xth XXXXXXXXXX 200X}}}

\noindent \textbf{\small{DOI: 10.1039/b000000x}}
\vspace{0.6cm}

\noindent \normalsize{Suspensions of copper nanowires are emerging as new electronic inks for next-generation flexible electronics. Using a novel surface acoustic wave driven extensional flow technique we are able to perform currently lacking analysis of these suspensions and their complex buffer. We observe extensional viscosities from 3 mPa$\cdot$s (1 mPa$\cdot$s shear viscosity) to 37.2 Pa$\cdot$s via changes in the suspension concentration, thus capturing low viscosities that have been historically very challenging to measure. These changes equate to an increase in the relative extensional viscosity of nearly 12,200 times at a volume fraction of just 0.027. We also find that interactions between the wires and the necessary polymer additive affect the rheology strongly. Polymer-induced elasticity shows a reduction as the buffer relaxation time falls from 819 to 59 \textmu s above a critical particle concentration. The results and technique presented here should aid in the future formulation of these promising nanowire suspensions and their efficient application as inks and coatings.}
\vspace{0.5cm}
 \end{@twocolumnfalse}
  ]

\section{Introduction}
\footnotetext{\dag~Electronic Supplementary Information (ESI) available. See DOI: 10.1039/b000000x/}

\footnotetext{\textit{$^{a}$~Department of Mechanical and Aerospace Engineering, Monash University, Clayton, Australia. Fax: 613 9905 1825; Tel: 613 9905 3480; E-mail: prabhakar.ranganathan@monash.edu}}
\footnotetext{\textit{$^{b}$~Department Chemical Engineering, Monash University, Clayton, Australia. }}
\footnotetext{\textit{$^{c}$~Micro/Nanophysics Research Laboratory, RMIT University, Melbourne, Australia. }}
\footnotetext{\textit{$^{d}$~Department of Mechanical and Aerospace Engineering, University of California-San Diego, San Diego, California, USA.}}

\begin{figure}[ht]
\includegraphics[width=8.7cm]{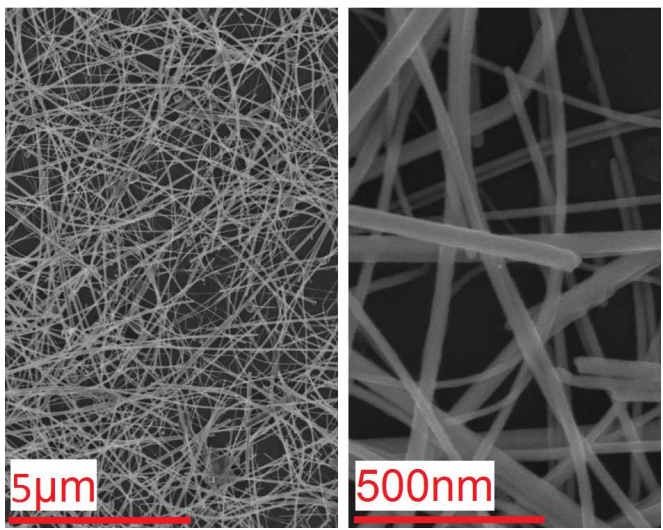}
\caption{\label{f:copperimg}Network-forming nature of long and thin CuNWs (approximately 27.6 nm in diameter, 62.36 \textmu m long, and possessing an aspect ratio of 2200 on average). A vast number of connections that is a result of annealing can be seen while a large area for light to penetrate is maintained.}
\end{figure}

Copper nanowires are novel building blocks in nanotechnology, exhibiting a wide range applications in electronics, field emission transistors, heat management, and electrochemical detectors \cite{bhanushali20141d}. Networks of thin and long copper nanowires (CuNWs) (see Fig. \ref{f:copperimg}) demonstrate very good electron transport in sparse concentrations after deposition and low temperature annealing (100 to 200\textsuperscript{o}C --- permitting the use of flexible plastic substrates). This leaves large spaces between the wires for the transmittance of light, spanning UV to infra-red wavelengths. Simultaneously, they also display high flexibly and stability \cite{Guo2013transparent}. These attributes make them appealing in the production of fully transparent and flexible conductive electrodes. Thus, they have been exploited in the development of devices such as light-emitting diodes, photovoltaic cells, stretchable conductors,  photodetectors, solar control windows, and touch screens, to name a few \cite{wu2013reversible}\textsuperscript{,}\cite{stewart2014solution}\textsuperscript{,}\cite{tang2014manufacturable}\textsuperscript{,}\cite{han2014fast}\textsuperscript{,}\cite{bhanushali20141d}\textsuperscript{,}\cite{tang2013ultralow}.

The utilisation of these particles is most likely to take the form of high-precision deposition or thin uniform coatings via a deposited suspension. Thus, there is a need for inks and coatings that are suitable for inkjetting, spraying, and drawing (among others) to be formulated, requiring a balance between the bulk fluid properties, method of delivery, and the target substrate. These application processes are dominated by extensional flow, complicating the development of suitable fluids for three reasons. First, these extensional processes predominantly require low-viscosity fluids to function\cite{magdassi2010chemistry}\textsuperscript{,}\cite{tracton2005coatings}\textsuperscript{,}\cite{fernando2000rheology}(the buffer used in this study was only 1 mPa$\cdot$s). Second, suspensions of CuNWs are likely to be highly non-Newtonian\cite{mewis1974rheological}\textsuperscript{,}\cite{batchelor1977effect}\textsuperscript{,}\cite{petrie1999rheology}. Third, these particles require polymer additives in order to stay suspended, lest they aggregate and settle. Hence, phenomena observed in polymer solutions in extensional flow such as long lived filaments and beads-on-a-string\cite{mun1998effects} must also be addressed. These challenges are compounded by the dearth of experimental extensional-flow methods that can handle low-viscosity fluids and quantify these surface-tension driven systems.

Inkjet technology generates extensional flows using low-viscosity fluids and it has been proposed that it can be used to characterise them\cite{basaran2013nonstandard}. However, inkjet systems have their limitations. Fromm (1984)\cite{fromm1984numerical} states that the critical properties that determine whether or not a particular fluid is suitable for inkjetting are the viscosity, density, surface tension, and the nozzle diameter. Given typical nozzle diameters of 10 and 40 \textmu m, ink viscosities of 2-20 mPa$\cdot$s and 5-50 mPa$\cdot$s are recommended, respectively, as evidenced by recent particle suspension ink work\cite{Shen2014nanoink}\textsuperscript{,}\cite{park2007direct}\textsuperscript{,}\cite{lee2008large}\textsuperscript{,}\cite{lawrie2013simple}\textsuperscript{,}\cite{grouchko2009formation}\textsuperscript{,}\cite{chen2010inkjet}\textsuperscript{,}\cite{Tao2013silvernano}. These viscosity ranges are not only narrow but also range below the lower limit of technologies like CaBER, an important consideration for characterising the complex low-viscosity buffers and dilute suspensions in this study. This point is further complicated here since the viscosity of these high-aspect ratio particle suspensions is anticipated to be highly sensitive to volume fraction, a potential hurdle in ink formulation when balancing conductivity against light transmittance. Further, it is recommended that the nozzle diameter is at least 100 times greater than the suspended particles\cite{magdassi2010chemistry}, limiting the range of sizes that can be handled by any one system - an issue likely complicated by these very high-aspect ratio particles, by which logic would potentially require nozzles of 6236 \textmu m. Further compounding the use of inkjets in the analysis of these long and thin wires are the complex flows and large shear rates generated within the nozzle itself that can be $2\times10^4$ to $2\times10^6$ s\textsuperscript{-1}\cite{derby2008bioprinting}, which may cause damage to NWs as they do to delicate proteins \cite{nishioka2004protein}\textsuperscript{,}\cite{derby2010inkjet}. Moreover, if thermal inkjet technology (as opposed to piezo) were used then the raised temperatures of around 300\textsuperscript{o}C exceed the annealing temperature for these particles, probably damaging them and possibly causing nozzle clogging. Exacerbating all this is the fact that every inkjet system has its own characteristics that produce system-specific results that can vary considerably. Though inkjet technology may be useful for delivering these suspensions in certain formulations, it is unlikely to be suitable for characterising them or complex buffers. Hence, finding a workable recipe to make inks from these NWs would otherwise be trial and error. 

Although accepted shear rheometry techniques have long been available\cite{dontula2005origins}, established extensional measurement methods are comparatively recent\cite{bazilevsky1990liquid}\textsuperscript{,}\cite{mckinley2002filament}. Capillary-breakup extensional rheometry (CaBER) involves rapidly extending a drop of sample fluid between two end-plates to a finite distance, creating a liquid filament with a critical aspect ratio that then thins due to surface tension. The thinning evolution is described by the balance of surface tension against inertial and viscous stresses caused by the extensional flow. Through this balance it is possible to extract the extensional viscosity by measuring the decay of the neck radius as a function of time \cite{kolte1997transient}. This technique has been used for highly viscous samples\cite{mckinley2000extract} but encounters problems with low-viscosity fluids. Rapid stoppage of the end plates causes vibrations to propagate through the fluid filament, leading to inertial instabilities which are not accounted for in the stress balance. These instabilities dissipate and can be ignored in higher viscosity samples or in samples with elasticity, where long-lived cylindrical polymeric filaments allow the use of a simple stress balance to extract the viscoelastic stresses\cite{entov1997effect}.

Recent advancements of the CaBER concept provide further access to low-viscosity fluids. The Cambridge Trimaster increases the end-plate retraction speed to overcome inertial instabilities and allow Newtonian and weakly elastic fluids of 10 mPa$\cdot$s to be tested \cite{vadillo2010evaluation}\textsuperscript{,}\cite{vadillo2012microsecond}. Another is the Slow-Retraction-Method\cite{campo2010slow}, in which a liquid filament is extended between two end-plates to a stable aspect-ratio, followed by a slow extension that pulls the filament beyond the critical aspect-ratio to initiate break-up. Electrowetting forces can also be used to pull a fluid filament to initiate capillary thinning\cite{EWOD2011}. A liquid to liquid annular flow microfluidic device has also been developed\cite{arratia2008polymeric}. The recently developed Rayleigh Ohnesorge Jetting Extensional Rheometry\cite{ROJER2015} technique involves stroboscopic imaging of a continuously flowing jet with controlled surface perturbations and has allowed access to very thin weakly viscoelastic fluids, though higher sample volumes would make such a device unsuitable for this study.

\begin{figure*}[ht]
\includegraphics[width=\textwidth]{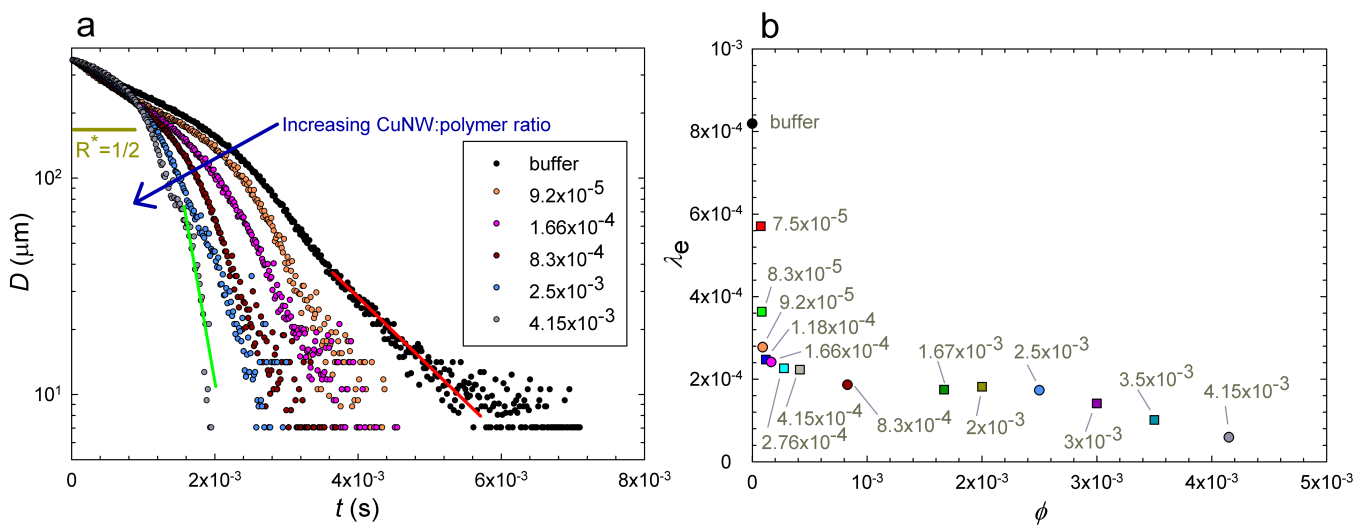}
\caption{\label{f:burel} (a) Filament break-up data clearly showing the reduction in polymeric effect with an increase in the CuNWs:polymer ratio. The red and green lines indicate the polymeric regions used calculate the relaxation times $\lambda_{\textnormal{e}}$ for the buffer and $\phi=$ $4.15\times10^{-3}$, respectively. Break-up above $R^{*}=1/2$ shows Newtonian-like behaviour, allowing extensional viscosity extraction as described in Section 2.1. The optical lower limit in these experiments occurs at 4 \textmu m. (b) Relaxation times showing a stark drop from 819 to 248 \textmu s at a CuNW volume fraction of $\phi=$ $1.18\times10^{-4}$, indicating a drop in the amount of free polymer in the solution. A more gently approached reduction to 59 \textmu s occurs at $4.15\times10^{-3}$. Grey annotations denote volume fractions.}
\end{figure*}

We use a novel variant of the CaBER technique that uses surface acoustic waves (SAW) to create and stabilise liquid bridges of very low viscosities; distinct advantages of this technique make it suitable for the development and analysis of these NW suspensions as inks and coatings. Firstly, it can process Newtonian fluids possessing shear viscosities from 27 Pa$\cdot$s down to 0.96 mPa$\cdot$s, giving it the range to encompass both viscous and characteristically difficult low-viscosity fluids\cite{bhattacharjee2011extensional}. Moreover, despite delivering large bulk mechanical energy the shear amplitudes created by high frequency SAW are so small that they can manipulate delicate particles, such as DNA, proteins and cells, without causing damage \cite{yeo2009ultrafast}\textsuperscript{,}\cite{li2009effect}\textsuperscript{,}\cite{Anushi2014}\textsuperscript{,}\cite{cortez2015pulmonary}. Further, SAWs have also been employed to pattern liquid crystals\cite{Liu2011} and nanotubes\cite{seemann2006alignment} without damage. Additionally, heating effects for the short bursts required to jet a droplet of water or glycerol have been determined to be less than 0.02 and 1\textsuperscript{o} C, respectively --- favourable for rheological analysis and well below the annealing temperatures of these metallic particles. Furthermore, there is no nozzle that can be clogged or that can restrict particle size. Finally, the fact that the technique only requires micro-droplets lends itself to the analysis of novel fluids such as biological samples that are difficult to produce in large volumes \cite{mcdonnell2015motility} (another microfluidic rheometric technique has also been recently proposed to handle small sample volumes\cite{erni2011microrheometry}). This is an important advantage since it is currently difficult to synthesise large quantities of CuNW particles.

Hence, we report changes in the extensional viscosity as a function of the particle concentration from dilute to highly concentrated and viscous suspensions. Additionally, the extensional properties of the polymer-doped aqueous buffer were also analysed, with unexpected phenomenological changes as the CuNW:polymer ratio was changed. The availability of this data and technique is anticipated to assist in the development and application of these promising particles. 

\section{Results and discussion}

\subsection{Polymeric buffer analysis}
\label{buffer}

Phase separation and settling due to depletion interaction has been reported in previous experiments with suspensions of particles in polymeric buffer\cite{bazilevskii2010sedimentation}. However, we observed no settling in our samples. This suggests that in our case polymers may absorb onto wires and serve to repel other coated wires, preventing aggregation and stabilising these suspensions. 

Capillary thinning of pure polymer solutions has been widely studied \cite{liang1994rheological}\textsuperscript{,}\cite{anna2001elasto}\textsuperscript{,}\cite{entov1997effect}\textsuperscript{,}\cite{clasen2006dilute}\textsuperscript{,}\cite{tirtaatmadja1993filament}\textsuperscript{,}\cite{prabhakar2006effect}. It is known that when a liquid bridge starts thinning, polymeric stresses are negligible, until polymer molecules have stretched sufficiently in the extensional flow at the necking plane. Molecular stretching leads to strain-hardening of the polymeric stresses and at some point, these stresses begin to dominate over the viscous stresses from the rest of the fluid. At that point, the decrease in filament radius changes from an approximately power-law like decay in low-viscosity solvents to an exponential decay. The relaxation time of this exponential decay is used as an indicator of the elasticity induced by the polymeric component.

We are interested in extracting the viscosity of the particle suspension alone without the interference from the polymeric component. A reduction in polymeric effects, quantified by the relaxation time $\lambda_{\textnormal{e}}$, is seen with an increase in the CuNW:polymer ratio (see Fig. \ref{f:burel}). The relaxation time drops from 819 \textmu s for pure buffer to 59 \textmu s at a volume fraction of $4.15\times10^{-3}$. We reason that as polymers bond to the wires they lose their ability to stretch under flow and the addition of more wires leaves fewer polymer molecules free to influence bulk fluid behaviour.  We then use the CuNW:polymer ratio at which $\lambda_{\textnormal{e}}$ is the smallest, while still maintaining the suspension, to minimise interference from the polymer, in either shear or extensional viscosity measurements.

\begin{figure}[ht]
\includegraphics[width=8.7cm]{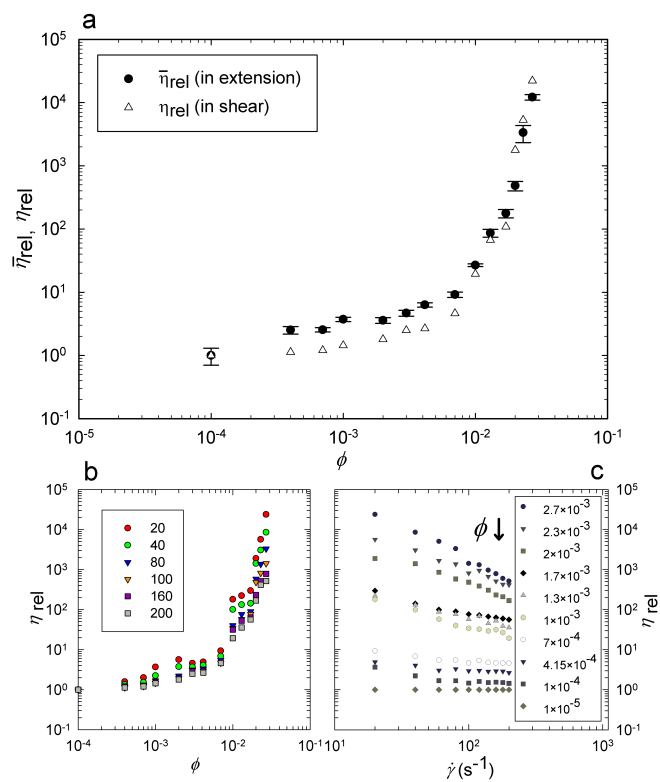}
\caption{\label{f:visc} (a) Relative extensional $\bar{\eta}_{rel}$ and shear viscosity $\eta_{rel}$ showing similar behavioural trends as a function of the volume fraction $\phi$. The shear viscosity values correspond to comparable extensional and shear strain rates. It should be noted that Sec.\ref{admier} describes how the extensional viscosities seen here were extracted. (b) Relative shear viscosity against volume fraction for shear rates $\dot{\gamma}$ of 20, 40, 80, 160, and 200 s\textsuperscript{-1}. (c) Relative shear viscosity against shear rate for different volume fractions, displaying pronounced shear thinning.}
\end{figure}

\subsection{Viscosity measurements}

Fig. \ref{f:visc}a shows the progressive change in the relative extensional viscosity $\bar{\eta}_{rel}=\bar{\eta}/3\eta_s$ with nanowire volume fraction, where $\eta_s = $1.01 mPa$\cdot$s is the shear viscosity of the buffer with the polymeric solute. A comparable growth in shear viscosities is also observed. It must be noted that each data point in Fig. \ref{f:visc}a is at a different strain-rate. In capillary break-up experiments, the strain-rate cannot be set \textit{a priori} but is self-determined by the fluid. In general, the average extensional strain-rate experienced by a sample prior to viscosity extraction decreases as volume fraction increases and viscosity increases. In shear rheometers on the other hand the strain rate can be pre-set. The shear data in Fig. \ref{f:visc}a were obtained at strain-rates comparable to the corresponding extensional flow experiment at the same volume fraction. The shear viscosity data obtained at fixed strain-rates are shown in Figs. \ref{f:visc}b and c. Since the polymeric component is expected to shear thin, and at any given volume fraction, the zero-shear-rate viscosity contribution of the polymer is small compared to the nanowire contribution (as demonstrated by the lack of significant exponential decay at small times during capillary break-up), the shear data in Figs. \ref{f:visc}b and c can also be considered to be free of polymeric interference.

Acute viscosity sensitivity to volume fraction is expected for these CuNW suspensions since these wires have the potential to directly interact with large numbers of particles along their lengths, and thus quick growth in the effects of inter-particle interaction relative to increased volume fraction was anticipated. In fact, the extensional viscosity more than doubled the buffer viscosity at a volume fraction of $1\times10^{-3}$. At $\phi=1.3\times10^{-2}$, an extensional viscosity of 265 mPa$\cdot$s was reached, well in excess of the typical inkjet system capability discussed in Section 1. Further, the relative extensional viscosity nearly rose to 12,200 times ($\bar{\eta}$ = 37.2 Pa$\cdot$s) at only $\phi=2.7\times10^{-2}$. The steep gradient at this point indicates that the maximum particle packing fraction was being approached even at this low volume fraction, which may be expected for such high aspect ratio particles\cite{parkhouse1995random}. 

It can be seen in Fig. \ref{f:visc}a that the shear viscosity overtakes the extensional viscosity in magnitude at about $\phi=1.7\times10^{-3}$. To our best knowledge, this has not been observed before. The reasons for this effect are not clear at present and may be attributed to the difference in the interplay between flow and particle orientation. The bare nanowires have an aspect-ratio of approximately A $=2\times10^{3}$. Even after accounting for an absorbed polymer layer, the aspect-ratio A $\approx O(10^{3})$. The volume fraction for the isotropic-nematic transition is expected to be around $1/A\approx 10^{-3}$. Most of the samples in Fig. \ref{f:visc} are well above this concentration. Extensional rheology of liquid-crystalline materials is relatively unexplored and the few studies thus far have focused on high-temperature liquid-crystalline polymer melts\cite{gotsis2000extensional}\textsuperscript{,}\cite{wilson1992transient}\textsuperscript{,}\cite{lin1988formation}. Extensional viscosities larger than shear viscosities have been observed in liquid-crystalline carbon-nanofibre suspensions, but at a single volume fraction\cite{xu2005shear}. CuNW suspensions could serve as model room-temperature liquid crystalline materials for more detailed explorations for understanding the effect of particle concentration on behaviour in extensional and shear flows. On the applied front, the data in Fig. \ref{f:visc}a show that to a large extent, shear viscosity can provide a good estimate of extensional viscosity for use in designing inkjet applications of CuNW suspensions.

\section{Experiments}

\subsection{Particle and buffer preparation}

CuNWs 25-30 nm in diameter and 50-65 \textmu m long were synthesised via a hydrothermal method\cite{jin2011shape}. 900 mg hexadecylamine (HDA)(Sigma-Aldrich) was dissolved in 50 ml water in a closed 100 ml screw cap borosilicate glass bottle, and magnetically stirred at 1000 rpm at 100\textsuperscript{o}C for 1 hour in an oil bath. To this, 100 mg CuCl$_{2}$2H$_{2}$O was added and stirred for another 30 minutes, after which 500 mg of $\alpha$ or $\beta$ glucose (Merck) was added and stirred at 400 rpm for 6 hours. The reaction solution takes on the appearance of a creamy white liquid with just the HDA after 1 hour. After the addition of CuCl$_{2}$2H$_{2}$O the solution turns sky blue after 30 minutes. Finally, after addition of glucose, the solution starts turning pale brown at first, then gradually dark brown, then reddish brown. The reaction is stopped after 6 hours and removed from the oil bath allowing it to cool for 10 minutes, after which it is transferred into 50 ml vials and centrifuged at 6500 rpm for 5 minutes. The CuNWs are collected at the bottom of the centrifuge tube, which can be easily separated by carefully decanting the waste supernatant. The pellet is carefully rinsed a few times with Milli-Q water at room temperature to remove any loosely bound HDA remaining.    

Hydroxypropyl cellulose (HPC)(Sigma Aldrich, 20 mesh, Mw = 1,000,000) was the polymer used as a dispersant to help suspend the CuNWs and prevent aggregation. A 2$\%$ wt solution of HPC in ethanol was made from of which a 0.6 ml aliquot was sperated into a test tube and diluted with Milli-Q water to make a 2.2 ml solution. This aqueous solution was used as the buffer, and the same amount of HPC solution was used with increasing volume fractions of CuNWs to find the optimum CuNW:polymer ratio through extensional analysis. 

\subsection{Acoustically-driven microfluidic extensional rheometry} \label{admier} 

\begin{figure}[ht]
\includegraphics[width=8.7cm]{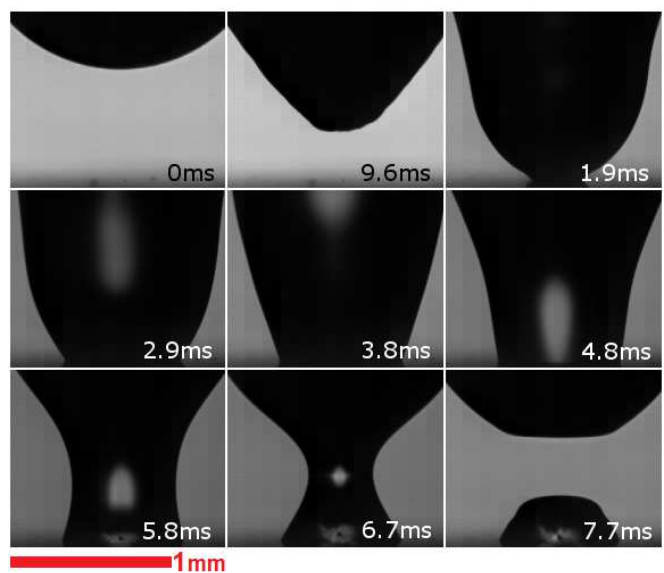}
\caption{\label{f:sawwatermont} A montage showing the SAW-actuated jetting, resulting from a 1.5 ms SAW burst, and subsequent surface tension-driven thinning of pure water at a temperature of 22\textsuperscript{o}C (0.96 mPa$\cdot$s).}
\end{figure}

We overcame the challenges discussed in Section 1 by using an experimental technique developed by Bhattacharjee \textit{et al.}, explained in detail elsewhere \cite{bhattacharjee2011extensional}\textsuperscript{,}\cite{mcdonnell2015motility}. Briefly, we used a 20 MHz waveform generator (33220A, Agilent) to trigger a second signal generator (WF1966, NF Corporation), which produces a 1.5 ms burst. The second generator outputs a sinusoidal signal to an RF power amplifier (411LA, ENI), resulting in a fixed amplitude and frequency signal of approximately 36.7 MHz that is then supplied to a piezoelectric SAW chip. This device comprises of a pair of curved inter-digitated transducers bonded to a piezoelectric lithium niobate substrate that focus Rayleigh waves towards a central target where a sessile droplet in placed (1 \textmu l; approximately 1 mm dia - Fig.\ref{f:sawwatermont}.) The SAW burst, focused beneath the droplet, concentrates considerable inertial energy within it, thus forcing a jet to form\cite{yeo2009ultrafast}. The SAW energy then stabilises the jet as it impacts onto an opposing surface and establishes a liquid bridge of a critical aspect ratio that will result in break-up. Upon relaxation of the SAW burst, the filament is allowed to thin under surface tension, the decay of which is recorded using a high speed camera (Photron SA5; 62000 fps; image size: 1.35 mm$\times$2.14 mm (192$\times$304 pixels) with a long-distance video microscope attachment (K2/SC, Infinity). The event is lit by a single LED lamp placed beyond the fluid filament. The radius of the neck in each image frame is extracted using standard image-analysis techniques. 

The method for extracting extensional viscosity for this technique is detailed elsewhere\cite{mcdonnell2015motility}. In short, the aforementioned difficulties of extracting data from thin cylindrical threads are addressed by beginning the measurement of filament thinning at a larger filament diameter, but not so large that initial transients affect data --- 50 pixels (0.352 $\pm$ 0.007 mm). A series of filament break-up data was acquired for a range of Newtonian fluids that represent viscosity standards, from low-viscosity to viscous samples. Break-up time and radius were re-scaled by the Rayleigh time-scale $t^{*}=t/\tau_{\textnormal{R}}$ and the initial radius $R^{*}=R/R_0$, respectively, where $\tau_{\textnormal{R}}\equiv\sqrt{\rho R_0^3/\sigma}$, $\rho$ is the density, $R_0$ is the initial radius, and $\sigma$ is the surface tension. When experimental conditions such as sample size were held constant, it was determined that the time $t^{*}_{1/2}$ for $R^{*}$ to decrease to a value of half was primarily a function of the Ohnesorge number Oh$\equiv\bar{\eta}/3\sqrt{\rho\sigma R_0}$, where $\bar{\eta}$ is the extensional viscosity, which is 3 times the shear viscosity $\eta$ for Newtonian fluids.  A calibration curve through $t^{*}_{1/2}$ - versus - Oh data for the Newtonian viscosity standards was obtained. With this calibration curve, the Oh and hence $\bar{\eta}$ of a test fluid sample with negligible elastic effects can be extracted from the observed $t^{*}_{1/2}$ measured. This method for extracting extensional viscosity is valid as long as viscoelastic effects due to the dissolved polymer are not significant, since the original calibration of dimensionless half-time against Ohnesorge number was only performed with Newtonian viscosity standards.

\subsection{Characterisation of polymeric filaments}

The addition of polymer to the suspension buffer necessitates quantification of viscoelastic properties. Polymer chains stretch under extensional flow, retarding the rate at which a fluid filament breaks up. These elastic stresses can dominate towards the later stages of break-up, resulting in long-lived filaments and can be characterised by a relaxation time $\lambda_{\textnormal{e}}= -t/3\ln(R(t)/R_0) $\cite{Bazilevsky1990}\textsuperscript{,}\cite{liang1994rheological}. Although in reality there may actually be more than one relaxation mode, it has been found that the slowest mode dominates in extensional flow, thus a single relaxation time is commonly used to characterise viscoelastic stress experimentally and theoretically \cite{entov1997effect}.

\subsection{Particle imaging} 

High resolution images of the CuNWs were captured using field-effect scanning electron microscopy (Nova NanoSEM 450, FEI, Oregon, USA). 

\subsection{Shear rheometry} 

A cone (60 mm/1$^o$, gap distance 0.041 mm) and plate rheometer was used to perform the shear viscosity analysis (HAAKE MARS III, Thermo Fisher Scientific, Massachusetts, USA). 

\section{Conclusions}

The capabilities of SAW-driven microfluidic extensional rheometery enabled the extensional analysis of NW suspensions, overcoming the challenges that low-viscosities, high-aspect ratio particles, and small sample volumes present.

The reduction of polymeric-induced elasticity with the addition of CuNWs, indicated by the relaxation time decreasing from 819 down to 59 \textmu s, suggests that polymer molecules bond to the CuNWs. Beyond a certain CuNW:polymer ratio, few free polymer molecules remain to influence bulk properties. Under these conditions, filament thinning was initially not significantly affected by the dissolved polymer, and Newtonian-like behaviour was observed at small strains. Hence, utilising this optimal ratio enabled the creation of sedimentation-free suspensions that did not exhibit complicating polymeric behaviours, thus allowing the observation of changes in viscosity that arise primarily as a function of CuNW volume fraction.

It was demonstrated that suspension viscosity of these long and thin wires are highly sensitive to volume fraction, with increases of 4 orders of magnitude in their relative extensional and shear viscosities, at a volume fraction of just $2.7\times10^{-2}$.

In summary, it is anticipated that the reported technique and viscosity data, particularly that of problematic low-viscosity fluids in extensional flow, will assist in the formulation of workable inks and coatings where controlling polymeric expression and extensional viscosity is key, thus expediting the application of these promising particles for the next generation of electronics.

\section{Acknowledgments}
The authors thank the Monash Center for Electron Microscopy facilities (MCEM) for the generous use of the FEI Nova NanoSEM 450 FESEM and the Victorian International Research Scholarship (VIRS).  

\appendix

\footnotesize{
\bibliography{mybib} 
\bibliographystyle{rsc} 
}
\begin{figure}[ht!]
\includegraphics[width=8.7cm]{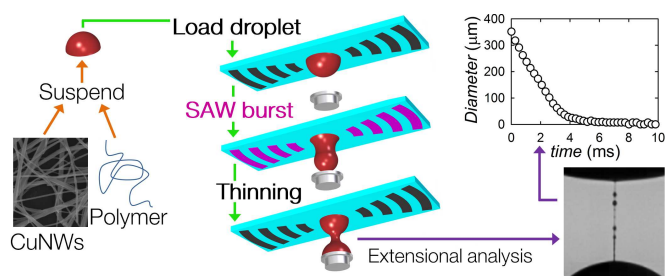}
\caption{\label{f:For Table of Contents Only} For Table of Contents Only. Novel experiments enabled the extraction of lacking extensional data on promising copper nanowire suspensions and their complex buffer, where low viscosities and long 1D particles pose significant challenges for alternative technologies.}
\end{figure}

\end{document}